# Strong spin-dephasing in a topological insulator-paramagnet heterostructure

Jason Lapano[1], Alessandro R. Mazza[1], Haoxiang Li[1], Debangshu Mukherjee[2], Elizabeth M. Skoropata[1], Jong Mok Ok[1], Hu Miao[1], Robert G. Moore[1], Thomas Z. Ward[1], Gyula Eres[1], Ho Nyung Lee[1], Matthew Brahlek[1*]

[1]Materials Science and Technology Division, Oak Ridge National Laboratory, Oak Ridge, TN, 37831, USA

[2]Center for Nanophase Materials Sciences, Oak Ridge National Laboratory, Oak Ridge, TN, 37831, USA

Correspondence should be addressed to *brahlekm@ornl.gov

**Abstract**: The interface between magnetic materials and topological insulators can drive the formation of exotic phases of matter and enable functionality through manipulation of the strong spin polarized transport. Here, we report that the spin-momentum-locked transport in the topological insulator $Bi_2Se_3$ is completely suppressed by scattering at a heterointerface with the kagome-lattice paramagnet, $Co_7Se_8$. $Bi_2Se_3$-$Co_7Se_8$-$Bi_2Se_3$ trilayer heterostructures were grown using molecular beam epitaxy. Magnetotransport measurements revealed a substantial suppression of the weak antilocalization effect for $Co_7Se_8$ at thicknesses as thin as a monolayer, indicating a strong dephasing mechanism. $Bi_{2-x}Co_xSe_3$ films, where Co is in a non-magnetic $3^+$ state, show weak antilocalization that survives to $x = 0.5$, which, in comparison with the heterostructures, suggests the unordered moments of the $Co^{2+}$ act as a far stronger dephasing element. This work highlights several important points regarding spin-polarized transport in topological insulator interfaces and how magnetic materials can be integrated with topological materials to realize both exotic phases as well as novel device functionality.

**Key Words**: Topological insulator, molecular beam epitaxy, heterostructures, weak antilocalization

Topology has entered condensed matter physics and materials science as a classification scheme that distinguishes materials based upon a single number, the topological invariant[1]. For the case of $Z_2$ topological insulators (TIs)[2], this invariant is determined by the ordering of the energy states of the band structure. At the boundary between 3-dimensional (3D) materials from different topological classes, new 2-dimensional (2D) singly degenerate energy states can emerge that possess linear Dirac-like dispersion and exhibit strong spin-momentum locked transport. These states can be modified by breaking time-reversal symmetry which gives many routes to control the surface state properties and electronic structure. Specifically, controllably opening a gap at the Dirac point can give rise to other exotic phases of topological matter, particularly, the quantized anomalous Hall phase[3–7] as well as states that where envisioned in the realm of particle physics, for example axion phases[8], which can be realized in condensed matter platforms. Due to these unique properties, the broad class of magnetic and topological materials are extremely promising for fundamental studies of exotic phases of matter as well as applications ranging from spintronics to quantum computing[9–12].

Numerous studies have focused on magnetically doping TIs to stabilize a ferromagnetic ground state[13–17], which has resulted in the observation of the quantum anomalous Hall phase[6]. Combining TIs and magnetic materials as heterostructures offers an alternate route over bulk doping as magnetic coupling between the ferromagnetic material and a TI can be achieved without excessive scattering off random magnetic impurities[18], as well as enabling the creation of distinct interfaces at the top and bottom surfaces, for example, which has been used to access axionic phases[19–21]. Specific questions remain regarding the fundamental scattering processes at interfaces between TIs and magnetic systems, particularly how the strong spin-momentum locking is modified in the presence of magnetic scattering centers and global magnetic ground states. Such questions are especially relevant for understanding interfaces with magnetic materials that are electrically insulating, for example yttrium iron garnet[22] and $Cr_2Ge_2Te_6$/$CrGeTe_3$[23,24], as well as metallic interfaces such as permalloy and Pt[25]. Here, we report that spin-momentum-locked transport in the TI $Bi_2Se_3$ is completely suppressed by scattering at the heterointerface with the kagome-lattice paramagnet, $Co_7Se_8$. Despite time-reversal symmetry being fully intact, these results highlight the sensitivity of the spin-polarized transport at interfaces with strong moments. This dramatic effect raises the question whether the topological properties are fundamentally compromised at the $Co_7Se_8$ interfaces as well as the 2 remote interfaces between the $Bi_2Se_3$ and the surface and substrate: to answer this we show that the topology is intact at the surface using *in situ* angle-resolved photo-emission spectroscopy (ARPES). This work highlights the nontrivial interplay among magnetic and topological materials, which can be used to realize both exotic phases as well as novel device functionality.

Key to this study is utilizing molecular beam epitaxy (MBE) growth to create atomically sharp interfaces with the TI, here, $Bi_2Se_3$ and the kagome paramagnetic material $Co_7Se_8$. $Bi_2Se_3$ is a prototypical topological insulator, having a relatively large band gap of 0.3 eV and a simple band structure with a single spin-polarized surface Dirac cone at the Γ point that is well separated from the 3D bulk bands[26–28]. As shown in the structural models in Fig. 1(a), the unit cell of $Bi_2Se_3$ is composed of three monolayers (ML) called quintuple layers (QL) that are roughly 1 nm thick. $Co_7Se_8$ has a monoclinic structure with a unit cell height of 0.52 nm, and 2 Co ML per unit cell. The $Co_7Se_8$ unit cell is characterized by an ordered-vacancy derivative of the hexagonal CoSe compound, in which 1/8 Co atoms are missing, creating alternating planes of close-packed hexagonal and kagome lattice structures[29], shown in Fig. 1(b). This yields Co with a mixed $2^+$ and $3^+$ valence state and a low-carrier density metal. Previous studies on bulk powder samples of $Co_7Se_8$ reported paramagnetism, and in nano-crystalline powders, trace amounts of Co or other Co-Se impurity phases can contribute to a weak ferromagnetic signature[29–31]. However, MBE growth typically results in highly uniform and homogeneous systems[32,33]. These respective materials enable the application of strong magnetic disorder on the TI without breaking time reversal symmetry, which allows a cleaner view of the mechanisms driving the spin polarized transport. Trilayer structures used here were composed of a 7 QL $Bi_2Se_3$ bottom layer, a variable thickness $Co_7Se_8$ layer ranging in number of monolayers from $n = 1$-23 ML, and a 7 QL $Bi_2Se_3$ top layer, as schematically shown in Fig. 1(c) along with a corresponding scanning transmission electron microscopy (STEM) image in Fig. 1(d). This structure was chosen since the bottom $Bi_2Se_3$ layer served as a necessary nucleation layer for $Co_7Se_8$, and the top $Bi_2Se_3$ layer preserved inversion symmetry between the $Co_7Se_8$ interfaces. Since the $Bi_2Se_3$ is in the regime where the bulk bands are occupied and the $Co_7Se_8$ is also metallic, both the top and bottom surface states can interact with the $Co_7Se_8$ through the free electrons in the bulk state of $Bi_2Se_3$. As such, this heterostructure enables probing both the nature of the spin-polarized transport as well as the topological character via ARPES on the top surface of the $Bi_2Se_3$.

Samples were grown using a home-built MBE system operating at a base pressure lower than $1\times10^{-9}$ Torr. Cobalt, bismuth, and selenium were all supplied via thermal effusion cells. The source temperature of each cell was adjusted prior to growth to give the desired flux that was measured by a quartz crystal microbalance. $Bi_2Se_3$ and $Co_7Se_8$ were both grown in a self-regulated manner, where the flux ratios between the Bi:Se and Co:Se were around 1:10. As such, the Se shutter remained open during the entire growth sequence, whereas the growth of $Bi_2Se_3$ and a $Co_7Se_8$ were, respectively, controlled by timing the opening of the shutters based on the individual fluxes. Samples were grown on 5×5 mm² $Al_2O_3$ (0001) substrates, which were treated prior to growth with an acetone and IPA ultrasonic bath to degrease the surface, followed by annealing in air at 1000 °C. In all cases, an initial 2 QL $Bi_2Se_3$ buffer layer was deposited at 150 °C to maintain a smooth interface[34]. The remainder of the heterostructures were grown at 200 °C, with the exception of the pure $Bi_2Se_3$ sample grown at 275 °C. Doped samples were grown by keeping the bismuth and selenium shutters open, while opening the Co shutter many times within a single monolayer to achieve homogeneous doping (see Fig. S1). X-ray diffraction measurements were performed on a Malvern Panalytical's X'Pert³ with a 4-circle goniometer using Cu $k_{\alpha 1}$ radiation. Transport measurements were performed in the van der Pauw geometry using pressed indium wires as contacts and measured in a Quantum Design Physical Property Measurement System down to a base temperature of 2 K. ARPES measurements were performed on a Scienta-Omicron ARPES system with a DA30-L electron analyzer, a helium lamp source with photon energy of 21.2 eV, and base temperature <10 K. The energy resolution of the ARPES measurement is 10 meV. Due to the difficulty in preserving the quality of the 2D materials during manual sample-preparation, STEM samples were thinned using a FEI Nova 200 focused ion beam (FIB), with the initial lift-out performed at an ion accelerating voltage of 30 kV, and final thinning performed at an accelerating voltage of 2 kV. Surface gallium ion implantation from the FIB thinning was removed by argon ion milling in a Fischione 1040 Nanomill, with 5 minutes per side at a voltage of 0.9 kV, which was followed by 1 minute of milling each side at 0.5 kV. The samples were subsequently imaged using a NION UltraSTEM 100 operating at an electron accelerating voltage of 100 kV, which was corrected for fifth order spherical aberrations. The images were collected using an annular dark field detector with the collection angles from 84-200 mrad. Images were collected with a pixel dwell time of 8 µs. Overall beam exposure was minimal due to observed sensitivity to beam-damage.

To understand the global structure of the trilayer heterostructures, XRD measurements were performed to probe crystallinity, morphology, and interfacial character. $2\theta$-$\theta$ scans of the parent materials, $Bi_2Se_3$, and $Co_7Se_8$, as well as the trilayer heterostructures for various $Co_7Se_8$ thicknesses are shown in Fig. 2(a). The $Bi_2Se_3$ sample was 14 nm, the $Co_7Se_8$ was 10 nm, and for the trilayer heterostructures the $Co_7Se_8$ thickness ranged from $n = 1\text{-}23$ ML. The $2\theta\text{-}\theta$ scan of $Bi_2Se_3$ shows the $003m$ series of peaks ($m$ is an integer), which are highlighted by square-symbols. The peaks due to the $Al_2O_3$ substrate are marked with asterisks. Further, for the $Bi_2Se_3$ peaks Laue oscillations due to coherent scattering off the top and bottom interfaces can be seen about the most intense peaks, which indicates the films are extremely flat. The $Co_7Se_8$ film shows only the 001 and 002 reflections, as marked by circles. The lack of Laue oscillations indicate that the films are slightly rougher or, since the intensity of the oscillations scale with the intensity of the reflection, that they are below the scan's detection threshold due to the relative weakness of $Co_7Se_8$ 001 and 002 peaks. The latter point is highlighted by comparing the 0012 reflection of $Bi_2Se_3$ at around $2\theta \approx 37°$, which is similar magnitude at the 002 peak of $Co_7Se_8$. Neither of these reflections show Laue oscillations, which indicates that the films are likely of similar flatness. The $2\theta\text{-}\theta$ scans of the trilayers samples, with $Co_7Se_8$ thickness ranging from $n = 1\text{-}23$ ML, reveal a complex array of features. For the single monolayer the data looks very similar to $Bi_2Se_3$, only with some of the reflections slightly distorted and with weak oscillations that are superimposed upon the main peaks. With increasing $Co_7Se_8$-thickness, the $Bi_2Se_3$ peaks are further distorted, and these unusual oscillations become more prominent. Moreover, at 4 ML, the $Co_7Se_8$ peaks clearly emerge and by 23 ML are relatively sharp, which indicate that a significant portion of scattering emanates solely from $Co_7Se_8$.

We further investigated the source of these intensity oscillations utilizing XRD simulations of the heterostructures for two plausible origins (see Supplementary Materials and Ref.[35] for additional details). For both cases, when the films are thick, interference should qualitatively occur if reflections from both the parent materials are close in $2\theta$. This occurs since the reflected X-rays can interfere constructively or destructively; away from these regions no coherent scattering should be seen. In the first scenario two pristine $Bi_2Se_3$ layers with sharp interfaces are separated by a thickness equivalent to $n$ monolayers of $Co_7Se_8$, representing an ideal structure, shown in Fig. 2(b) for the $n=8$ condition labelled "ideal 8 ML". This accounts for interference off the various interfaces, and clearly reproduces the experimental data. In the second scenario, we assume some interdiffusion of Co into the $Bi_2Se_3$ and perhaps Bi into the $Co_7Se_8$ which will likely occur asymmetrically if the $Bi_2Se_3/Co_7Se_8$ interface is different than the subsequent $Co_7Se_8/Bi_2Se_3$ interface[33,36]. This would be accompanied by a modification of the out-of-plane lattice parameter in the various layers[37], thus creating, for example, peak broadening or multiple $Bi_2Se_3$ reflections very close in $2\theta$ depending on the degree of lattice distortion and cobalt interdiffusion. However, interference should only occur at lower $2\theta$ where the overlap of the reflections would be maximum and would give way to a splitting of the peaks at higher $2\theta$. This is shown in the simulation in Fig. 2(b) labeled "slightly mixed" and "heavily mixed", where the intensity is calculated for a bilayer structure composed of two $Bi_2Se_3$ of slightly different out-of-plane lattice parameters. For the lightly mixed case, a small (1%) variation in the lattice constant from the bulk $Bi_2Se_3$ reproduced the reflections at high $2\theta$ values, but not the interference pattern at lower $2\theta$. For the heavily mixed sample which had a larger (5%) variation in the lattice constant, there is some interference at low $2\theta$ values but a large degree of peak splitting at high $2\theta$. Moreover, X-ray photoelectron spectroscopy (XPS) was carried out on the trilayer with $n = 8$ ML and doped samples and is shown in Fig. S2. No cobalt signature could be seen for the trilayer sample, confirming that there is minimal diffusion of Co into the $Bi_2Se_3$ layers and that the surface is free of Co impurities. As such, this scenario can be eliminated. Considering this and returning to the experimental data in Fig. 2(a), comparing the trilayer structures to the simulated data reveals only the ideal case closely matches the experimental data at both high and low $2\theta$ values, confirming the global structural quality of the trilayer heterostructures.

Now that the structural quality has been confirmed, we move on to magnetotransport measurements. Conductance versus magnetic field is shown in Fig. 3(a) for the trilayer samples in comparison to the 14 nm $Bi_2Se_3$ (0 ML) film measured at 2 K. With increasing magnetic field, the conductance for the $Bi_2Se_3$ sample shows a sharp drop with increasing field; at higher fields the conductance transitions to a weak $B^2$ dependence, where $B$ is the magnetic field, which is characteristic of the free-electron response. For the trilayer heterostructures, the sharp drop in the conductance is substantially suppressed even for samples with $Co_7Se_8$ thickness as thin as a single monolayer, and by 4 ML the conductance exhibits solely a $B^2$ dependence. Transport in TIs in low magnetic field is characterized by a sharp cusp-like feature centered around $B = 0$ T[38]. The origin of this feature, called weak antilocalization (WAL), is due to the magnetic field suppressing coherent backscattering. In TIs specifically, and, more generally, 2D systems with strong spin-orbit coupling, WAL is due to a reduction of the probability to backscatter, which is driven by the accumulation of phases of opposite sign for clockwise and counter-clockwise backscattering loops[39]. Application of a small magnetic field, therefore, suppresses this effect, which is manifest as a sharp drop in conductance. The WAL effect can be quantified by fitting the experimental data using the Hikami-Larkin-Nagaoka (HLN) model for the change in conductance,

$$\Delta G(B) = \alpha \left(\frac{e^2}{2\pi h}\right)\left[\ln\left(\frac{B_\emptyset}{B}\right) - \Psi\left(\frac{1}{2} + \frac{B_\emptyset}{B}\right)\right] \quad (1)$$

in which $\Psi(x)$ is the digamma function, $h$ is Planck's constant, $e$ is the electron charge, and the two free parameters are $\alpha$ which quantifies the number of 2D channels (1 per channel; this has been rescaled by $1/(2\pi)$ relative to the original HLN formulation), and $B_\emptyset$ is the dephasing magnetic field[40]. The dephasing magnetic field is related to the phase coherence length $l_\emptyset$ by the following equation.

$$B_\emptyset = \frac{\hbar}{4el_\emptyset^2} \quad (2)$$

The data was fit using the HLN model with an additional $B^2$ term to account for the free electron response. The fits are shown in Fig. 3(a) as solid lines for the $Bi_2Se_3$ and the trilayer with $n = 1$. The resulting $l_\emptyset$ and α are plotted versus $Co_7Se_8$ thickness in Fig. 3(b) and (c), respectively. For larger thicknesses, the low-field kink was absent, and, therefore, we took both $l_\emptyset$ and $\alpha$ to be zero indicating that there is no spin-polarized transport. The complete quenching of this effect is surprising since the transport for TIs should be immune from such scattering events so long as time reversal symmetry is intact. This immunity is demonstrated by the magnetoconductance data shown in Fig. 3(d). This data was taken from $Bi_{2-x}Co_xSe_3$ where $x$ ranged from 0.07 to 0.50. For these films it is assumed that Co takes a non-magnetic $3^+$ valence state and predominately replaces $Bi^{3+}$. Over this range the WAL effect can be seen clearly to survive to the highest doping range as the cusp in the conductance in the low field regime is maintained. Fitting to the HLN function shows that $\alpha$ is nominally constant and $l_\phi$ drops relative to the undoped materials.

To understand the data for both the trilayer structures and doped samples it is instructive to compare length scales for the scattering processes. For the trilayer samples scattering off the $Co_7Se_8$ layer provides complete dephasing, thus, the minimum length scale is then set by the $Bi_2Se_3$ thickness. Physically this implies that all backscattering loops that contribute to WAL effect contain at least one scattering event off the $Co_7Se_8$ layer. For $Bi_2Se_3$, where both the bulk and the surface states contribute to the WAL effect, scattering off the surfaces does not provide a dephasing mechanism, since WAL is observed from the thick limit of hundreds of nanometers down to the thin limit where surface scattering dominates the resistance[38,41,42]; moreover, in $Bi_2Se_3/(Bi_{1-x}In_x)_2Se_3/Bi_2Se_3$ trilayer heterostructures, there a strong dependence on the doping levels and thickness of the $(Bi_{1-x}In_x)_2Se_3$[43]. In contrast to this, in $Bi_{2-x}Co_xSe_3$, the length scale is approximately set by the average spacing of the dopants, which is of the order of a few nanometers for $x \approx 0.1$. Therefore, scattering due to Co defects should occur at a significantly higher rate than scattering off the $Co_7Se_8$ interfaces in the trilayer heterostructures. Yet, WAL is clearly observed for the highest doped samples. It is emphasized that the Co should be in a $3^+$ valence state with no net moment. This highlights the extreme sensitivity of the spin-polarized transport processes to magnetic defects.

As the spin-polarized transport is fully suppressed, the question arises whether the trilayer system is topological. To answer this question *in situ* ARPES was performed on a trilayer sample with a $Co_7Se_8$ thickness of 8 ML at low temperature (<10 K). The spectrum for this sample is shown on the left of Fig. 4 (a). From this the linearly dispersive topological surface band is visible. This, and the degenerate Dirac point (DP), can be more clearly seen by processing the spectrum with the curvature method described in Ref. [44], as shown on the right-hand side of Fig. 4(a) and the equipotential map at various energies shown in Fig. 4(b). To further illustrate the gapless nature of the DP, we plot the stacking energy distribution curves (EDCs) in Fig. 4(c). The red curve indicates the EDC at $k_x = 0$, which shows a single peak at ~300 meV that corresponds to the DP position in the 2D spectrum in Fig. 4(a). However, the ARPES measurement only probes the top surface of the trilayer heterostructure. The electronic structure of the interface between $Co_7Se_8$ and $Bi_2Se_3$ remains uncertain. This, however, motivates the future question regarding the nature of the states that form at the interfaces of $Co_7Se_8$. The strong magnetic moments of the Co will certainly interfere with the spin polarized states, but strong hybridization among the $Co_7Se_8$ states and the topological surface states may also inhibit its formation entirely. Since this interface is well below the escape depth for photoelectrons used here, this question will have to be addressed in the future in combination with first principles calculations.

To conclude, we have shown that spin-polarized transport is completely suppressed in the topological material $Bi_2Se_3$ by embedding an epitaxial layer of a kagome-paramagnetic material, $Co_7Se_8$, using MBE growth. Scattering off the magnetic $Co^{2+}$ in the $Co_7Se_8$ is the primary source of the strong spin dephasing. This is in stark contrast to $Bi_{2-x}Co_xSe_3$ where spin polarized transport is found to survive the non-magnetic disorder due to $Co^{3+}$ even in doping levels as high as $x = 0.5$. *In situ* ARPES measurements show that, despite the absence of spin polarized transport, the topological band structure on the top surface remains intact. Taken with a broader prospective, the current measurements have deep implications. One of the biggest impacts comes from device functionality that relies on spin-polarization. Topological materials are supposed to be immune to disorder that obeys time reversal symmetry. Although $Co_7Se_8$ is not magnetic, it does have a net moment which causes complete spin dephasing of the charge carriers. As such, this demonstrates that in a device the materials need to be carefully chosen to avoid anything with a net moment—specifically, paramagnetic substrates with a net moment should be avoided for the growth of topological materials. Alternatively, if control over spin dephasing is desired, strong moment materials such as $Co_7Se_8$ can be incorporated epitaxially with TIs. Further questions regarding the source of dephasing in topological systems necessitates understanding and quantifying the dependence of the scattering processes on the details of the interfaces. As an example, metallic materials with a net moment, like $Co_7Se_8$, may be more susceptible to suppressing spin-polarized transport processes than an insulating material with net moments. As such, proposals for realizing novel topological phases necessitate interfacing magnetic metals, insulators, or both, thus highlighting the significance of these questions. This work brings about a broader view of spin polarized transport processes in topological materials, which is significant for realizing novel phases of matter at topological interfaces and for engineering topological devices.

## Supplementary Materials

See supplementary materials for additional X-ray diffraction data, details regarding the simulations, and X-rap photoemission spectroscopy.

## Acknowledgements

This work was supported by the U.S. Department of Energy (DOE), Office of Science, Basic Energy Sciences, Materials Sciences and Engineering Division (transport, structural characterization, and MBE growth) and by the Laboratory Directed Research and Development Program of Oak Ridge National Laboratory, managed by UT-Battelle, LLC, for the U.S. DOE under Contract No. DE-AC05-00OR22725 (ARPES measurements). The electron microscopy work was conducted as a user project at the Center for Nanophase Materials Sciences, which is a U.S. DOE Office of Science User Facility.

## AIP Publishing data sharing policy

The data that support the findings of this study are available from the corresponding author upon reasonable request.

# Figures

**Fig. 1. Materials and experimental schematic.** (a) Crystal structure of TI $Bi_2Se_3$ showing the layered hexagonal structure consisting of 3 repeated quintuple layer (QL) units of Se-Bi-Se-Bi-Se, three of which make the unit cell. (b) $Co_7Se_8$ is characterized by an ordered vacancy structure forming a kagome network of Co atoms in layer 1 and a hexagonal structure in layer 2. This leads to a mixed valence of $2^+$ and $3^+$ states, which are shown at the bottom decorating the layers as red arrows and blue circles, respectively. The structure is formally monoclinic, however the $\langle 1\bar{1}0 \rangle$ can be thought of as the pseudohexagonal $\langle 010 \rangle$ direction, showing the hexagonal arrangement between the two unit cells. (c-d) Schematic of the trilayer structure (c) and accompanying scanning high-angle annular dark field (HAADF) scanning transmission electron microscopy (STEM) image (d) for a $Bi_2Se_3$ (7 QL)/$Co_7Se_8$ (23 ML)/$Bi_2Se_3$ (7 QL) trilayer sample.

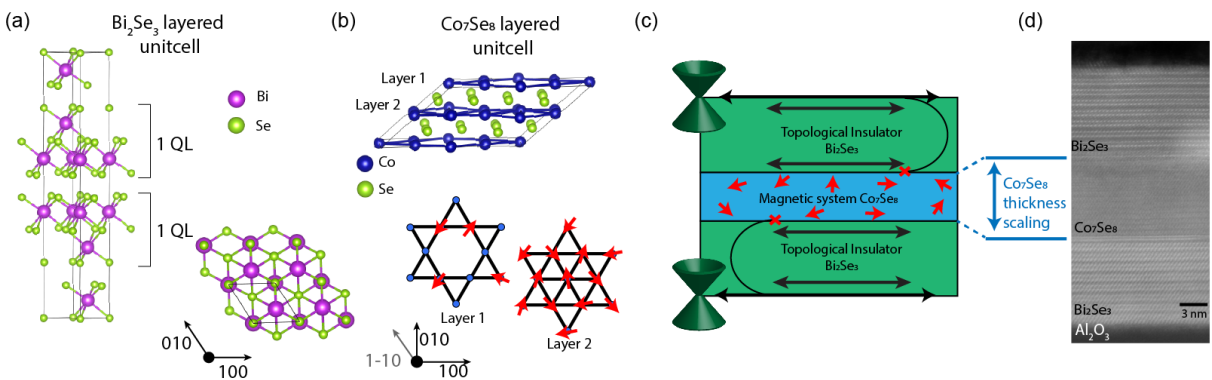

**Fig. 2. X-ray diffraction (XRD) measurements and simulations of trilayer heterostructures and parent materials.** (a) XRD $2\theta$-$\theta$ scans of the parent materials $Bi_2Se_3$ and $Co_7Se_8$ as well as trilayer heterostructures with $Co_7Se_8$ thickness ranging from $n$ = 1-23 monolayers (ML). The $Bi_2Se_3$ peaks are marked by square-symbols, $Co_7Se_8$ peaks are marked by circles, and the $Al_2O_3$ peaks are marked by asterisks. (b) Results from simulated structures including single-layer $Bi_2Se_3$, bi-layer $Bi_2Se_3$ with differing $c$-axis lattice parameters labeled "lightly mixed" and "heavily mixed" (the expected splitting at large $2\theta$ is labeled "$\Delta 2\theta$",) as well as a $Bi_2Se_3$ heterostructure with spacing equivalent to 8 ML of $Co_7Se_8$ labeled "ideal 8 ML" (see Supplement Materials). X-ray diffraction of the doped samples can be found in Fig. S1.

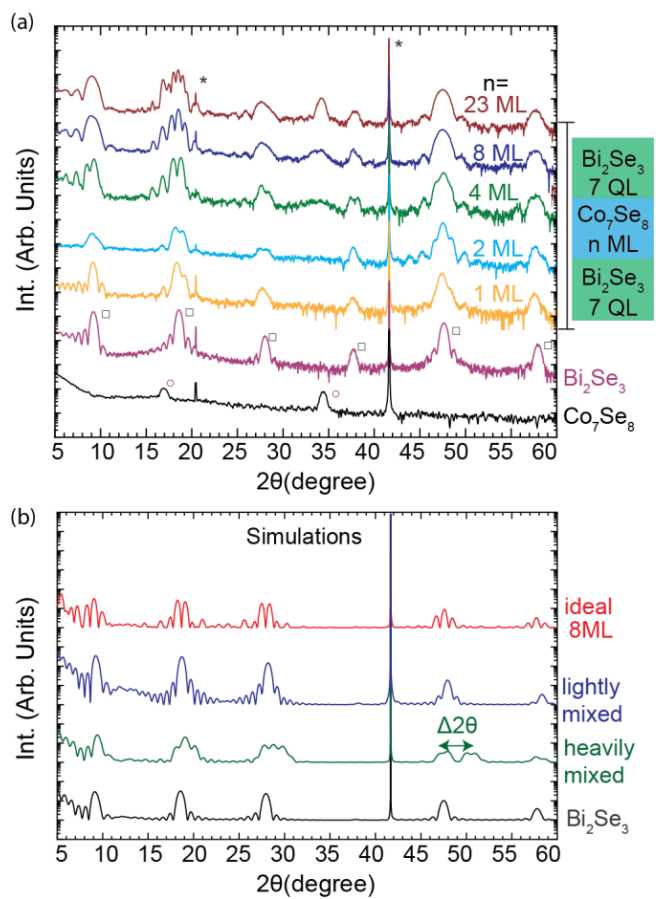

**Fig. 3. Conductance versus magnetic field for trilayer heterostructures and doped samples.** (a,d) Change in conductance versus magnetic field for the trilayer heterostructures (a) and for $Bi_{2-x}Co_xSe_3$ (d) where the data is shown as symbols and fit is the solid lines. Curves are offset for clarity. (b-c,e-f) Extracted α (number of conductance channels) and $l_\phi$ (phase coherence length) parameters from the HLN model are plotted for the trilayers (b-c) and $Bi_{2-x}Co_xSe_3$ (e-f), respectively. The single $Bi_2Se_3$ layer is the black square symbol, the trilayer are blue triangles, and the $Bi_{2-x}Co_xSe_3$ are the red circles. The dashed lines are guides-to-the-eye.

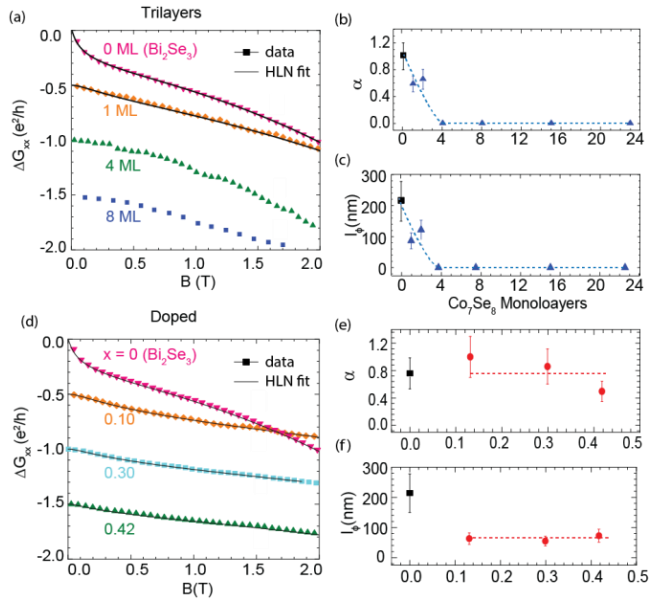

**Fig. 4. Angle-resolved photo-emission spectroscopy (ARPES) spectrum and equipotential maps for a trilayer heterostructure with 8 ML $Co_7Se_8$ thickness.** (a) ARPES spectrum cut across the $\Gamma$ point along the $k_x$ momentum direction. Left panel is the raw APRES spectrum while the right panel is the 2$^{nd}$ derivative of the spectrum to enhance the peak intensity of the dispersion[44]. (b) Equipotential maps taken above and below the Dirac point, which is 300 meV below the Fermi level. (c) Stacking energy distribution curves (EDCs) of the ARPES spectrum in panel (a). The red curve indicates the EDC across the Dirac point (DP) at $k_x=0$.

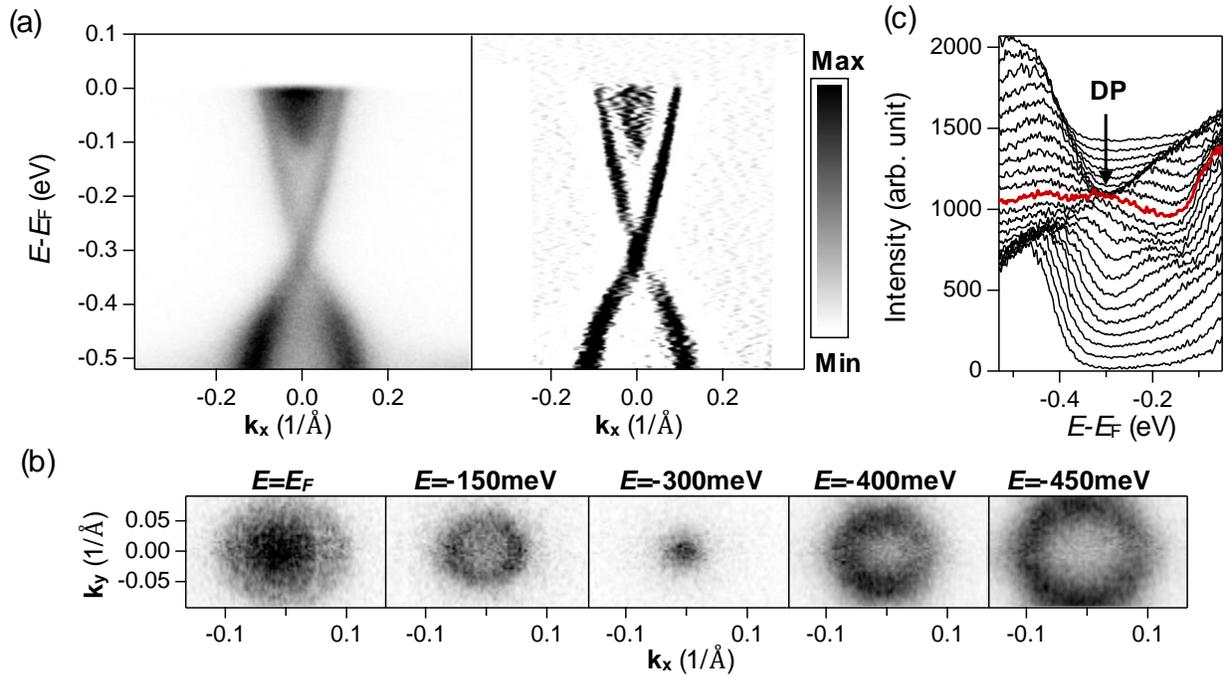

## Supplemental Materials:

# Strong spin-dephasing in a topological insulator-paramagnet heterostructure

Jason Lapano[1], Alessandro R. Mazza[1], Haoxiang Li[1], Debangshu Mukhergee[2], Elizabeth M. Skoropata[1], Jong Mok Ok[1], Hu Miao[1], Robert G. Moore[1], Gyula Eres[1], Zac Ward[1], Ho Nyung Lee[1], Matthew Brahlek[1*]

[1]Materials Science and Technology Division, Oak Ridge National Laboratory, Oak Ridge, TN, 37831, USA

[2]Center for Nanophase Materials Sciences, Oak Ridge National Laboratory, Oak Ridge, TN, 37831, USA

Correspondence should be addressed to *brahlekm@ornl.gov

**Fig. S1.** X-ray diffraction (XRD) measurements of doped $Bi_{2-x}Co_xSe_3$ series. The asterisk (*) indicates the peaks due to the $Al_2O_3$ substrate.

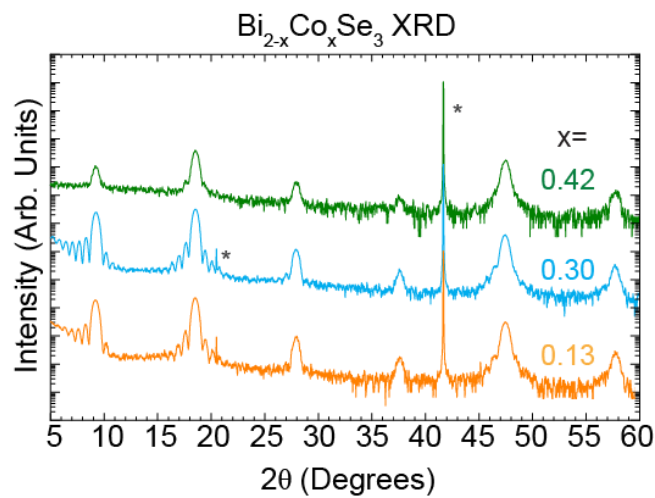

**X-ray diffraction simulations of trilayer samples.**

Insight into the structure of trilayer films was obtained by modeling of the specular Bragg reflections[1]. The model was written in terms of out-of-plane scattering vector $q_z = 4\pi\sin(2\theta/2)/\lambda$ where for the instrument used $\lambda = 1.5406$ Å. The X-ray diffraction from a thin-film epitaxial sample is given by the square modulus of the sum of amplitudes from the individual layers. The specular intensity, $I_{Spec}$, is calculated using a kinematic model as a function of $q_z$ and is given by

$$I_{Spec}(q_z) = \frac{S}{q_z^2}\left|\frac{F_{Sub}(q_z)}{1-e^{-iq_z c_{Sub}}} + A_{BiSe,1}(q_z) + e^{iq_z(t_{BiSe,1}+t_{CoSe})}A_{BiSe,2}(q_z)\right|^2, \quad (1)$$

$$F_{Sub}(q_z) = \sum_{j=1}^{6}\rho_O e^{iq_z(\frac{1}{12}+\frac{1}{6}j)c_{Sub}} + \rho_{Al}e^{iq_z(0.0188+\frac{1}{6}j)c_{Sub}}, \quad (2)$$

and

$$A_{BiSe,m}(q_z) = \sum_{j=0}^{n-1}\rho_{Bi}\left(e^{iq_z(j+0.18)d_{BiSe,m}} + e^{iq_z(j+0.56)d_{BiSe,m}}\right)e^{-\sigma_{j,Bi}^2 q_z^2} + \rho_{Se}\left(e^{iq_z(j)d_{BiSe,m}}\right.$$
$$\left. + e^{iq_z(j+037)d_{BiSe,m}} + e^{iq_z(j+0.735)d_{BiSe,m}}\right)e^{-\sigma_{j,Se}^2 q_z^2} \quad (3)$$

where the full structure factor of the sapphire substrate is $F_{Sub}(q_z)$ and for the Bi$_2$Se$_3$ layers $A_{BiSe,m}(q_z)$, which is indexed by $m = 1$ for the bottom layer and $m = 2$ for the top layer. The parameters used are as follows:

In eq. (1) $S$ is an overall scale factor, the term in the denominator of the for Al$_2$O$_3$ accounts for the surface truncation, and, in the third term, the phase term, $e^{iq_z(t_{BiSe,1}+t_{CoSe})}$ models the spatial separation among the two Bi$_2$Se$_3$ layers where $t_{BiSe,1}$ is the total thickness of the first Bi$_2$Se$_3$ layer and $t_{CoSe}$ is the thickness of the equivalent Co$_7$Se$_8$ layer. In eq. (2) for Al$_2$O$_3$ $c_{Sub} = 12.991$ Å is the lattice parameter along the $c$-axis and the terms in the parenthesis in the exponent index the atomic positions within the unit cell. For eq. (2-3) $\rho_i$ is the atomic weight as an approximation to the atomic form factor. In eq (3), $d_{BiSe} = 9.526$ Å, $n = 7$ is the number of unit cells of Bi$_2$Se$_3$, and again the terms in the parenthesis in the exponent index the atomic positions within the unit cell. For the simulation of the ideal structure the thickness of the Co$_7$Se$_8$ is 8 ML which is about 41 Å. For the lightly mixed structure $d_{BiSe,1}=d_{BiSe}$, and $d_{Bi2Se,2}=0.99\times d_{BiSe}$, and $t_{CoSe} = 0$ Å$^2$, while the heavily mixed sample is structure $d_{Bi2Se,2}=0.95\times d_{BiSe}$ $^2$. Finally, the exponential $e^{-\sigma_{j,Bi/Se}^2 q_z^2}$ is the Debye-Waller term that was included to account for vertical fluctuations in the atomic positions that appears as a fall off of the intensity with increasing $q_z$, and only changes the relative intensities of the reflections. The semiempirical parameters $\sigma_{j,Bi} = 0.001$ and $\sigma_{j,Se} = 0.01$ is related to the root-mean-square vertical displacements of atoms[1,3].

**Fig. S2** Surface compositional analysis by X-ray photoelectron spectroscopy (XPS) performed on the 8 ML trilayer heterostructure (bottom black curve), doped sample with $x = 0.7$ (middle red curve), and $Co_7Se_8$ thin film (top blue curve). Cobalt is clearly visible on the surface of the doped sample and while no cobalt peak is observed for the trilayer structure, indicating the $Bi_2Se_3$ surface is free of Co impurities.

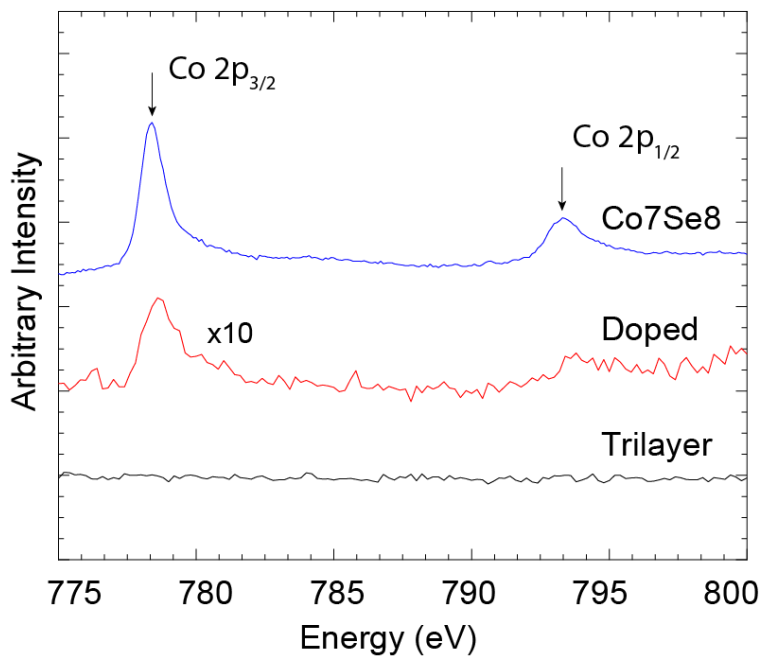